\documentclass{aa}  

\usepackage{graphicx}
\usepackage{txfonts}
\usepackage{natbib}
\newcommand{\bz}{\ensuremath{\langle B_z\rangle}}
\newcommand{\bs}{\ensuremath{\langle B \rangle}}
\newcommand{\nz}{\ensuremath{\langle N_z\rangle}}
\newcommand{\kms}{\ensuremath{\mathrm{km\,s}^{-1}}}
\newcommand{\vsi}{\ensuremath{v \sin i}}
\newcommand{\te}{\ensuremath{T_{\mathrm{eff}}}}
\newcommand{\eri}{{\rm 40~Eri~B}}
\begin{document}

   \title{A novel and sensitive method for measuring  very
     weak magnetic fields of DA white dwarfs}

   \subtitle{A search for a magnetic field at the 250~G level in 40~Eri~B}

   \titlerunning{Measuring very weak magnetic fields in DA white dwarfs}


   \author{J. D. Landstreet \inst{1,2} 
     \and S. Bagnulo\inst{1} 
     \and G. G. Valyavin\inst{3} 
     \and D. Gadelshin\inst{3} 
     \and A. J. Martin\inst{1,4}
     \and G. Galazutdinov\inst{5,6,3} 
     \and E. Semenko\inst{3}\fnmsep\thanks{Based in part on observations obtained at
       the Canada-France-Hawaii Telescope (CFHT) which is operated by
       the National Research Council of Canada, the Institut National
       des Sciences de l'Univers of the Centre National de la
       Recherche Scientifique of France, and the University of
       Hawaii.}
} 

   \institute{Armagh Observatory, College Hill, Armagh, BT61 9DG, 
              Northern Ireland, UK
              \email{jls@arm.ac.uk;sba@arm.ac.uk}
         \and
              Department of Physics \& Astronomy, University of 
              Western Ontario, London, Ontario N6A 3K7, Canada\\
         \and
              Special Astrophysical Observatory, Nizhnij Arkhyz, 
              Zelenchukskiy Region, 369167 Karachai-Cherkessian Republic, 
              Russia\\
         \and
              Astrophysics Group, Keele University, Keele, Staffordshire, 
              ST5 5BG, UK \\
         \and Instituto de Astronomia, Universidad Catolica del Norte, 
              Av. Angamos 0610, Antofagasta Chile\\
         \and Pulkovo Observatory, Pulkovskoe Shosse 65, 
              Saint-Petersburg 196140, Russia\\
             }

   \date{Received September 15, 1996; accepted March 16, 1997}

 
  \abstract
   {Searches for magnetic fields in white dwarfs have clarified  both the
     frequency of  occurrence and the global structure of the fields
     found down to field strengths of the order of 500~kG. Below this
     level, the situation is still very unclear.}
   {We are engaged in a project to find and study the weakest magnetic
     fields that are detectable in white dwarfs, in order to
     empirically determine how the frequency of occurrence and the
     structure of fields present changes with field strength. In this
     paper we report the successful testing of a very sensitive method
     of longitudinal field detection in DA white dwarfs. We use this
     method to carry out an extremely sensitive search for magnetism
     in the bright white dwarf \eri.}
   {The method of field measurement we use is to measure, at high
     spectral resolution, the polarisation signal $V/I$
     of the narrow non-LTE line core in H$\alpha$ in DA stars. This
     small feature provides a much higher amplitude polarisation
     signal than the broad Balmer line wings.  We test the usefulness
     of this technique by searching for a weak magnetic field in
     \eri. }
   {One hour of observation of $I$ and $V$ Stokes components of the
     white dwarf \eri\ using ESPaDOnS at the CFHT is found to provide
     a standard error of measurement of the mean longitudinal magnetic
     field \bz\ of about 85~G. This is the smallest standard error of
     field measurement ever obtained for a white dwarf. The
     non-detections obtained are generally consistent with slightly
     less accurate measurements of \eri\ obtained with ISIS at the
     WHT and the Main Stellar Spectrograph at SAO, in order to
     provide comparison standards for the new method. These further
     measurements allow us to make a quantitative comparison of the
     relative efficiencies of low-resolution spectropolarimetery
     (using most or all of the Balmer lines) with the new method
     (using only the core of H$\alpha$).  }
   {The new method of field detection reaches the level of sensitivity
     that was expected. It appears that for suitable DA stars, about
     the same field uncertainties can be reached with ESPaDOnS on the
     CFHT, in a given integration time, as with FORS on an 8-m
     telescope, and uncertainties are a factor of two better than
     with low-resolution spectropolarimetry with other 4--6-m class
     telescopes. However, even with this extraordinary sensitivity,
     there is no clear indication of the presence of any magnetic
     field in 40~Eri~B above the level of about 250~G. }


   \maketitle
%

\section{Introduction}

Megagauss magnetic fields were first discovered in white dwarfs in the 
1970s by detection of broadband circular (and soon afterwards linear)
polarisation \citep{Kempetal70,AngeLand70b,AngeLand71}. It was already
understood that such huge fields, and indeed much weaker ones, could
be detected by observing the Zeeman effect, for example in Balmer
lines \citep{AngeLand70a}, and since the early 1970s the overwhelming
majority of magnetic white dwarfs have been discovered using this
effect, which in practice is sensitive to fields from some kG to
hundreds of MG in white dwarfs
\citep{Land92,VanLandetal05,Kawketal07}. At present many hundreds of
magnetic white dwarfs are known as a result of searches through the
thousands of white dwarf spectra obtained by the Sloan Digital Sky
Survey \citep{Kepletal13}. It appears that magnetic fields above $\sim
1$~MG are present in roughly 10\% of all white dwarfs
\citep{Liebetal03}.

The fields of the magnetic white dwarfs are usually measured using the
splitting of spectral lines, which provides an estimate of the modulus
of the field averaged over the visible stellar hemisphere \bs. The
fields can also be measured using the (usually circular) polarisation
signature produced by the Zeeman effect, which can be used to estimate
the mean line-of-sight component of the magnetic field, averaged over
the visible hemisphere \bz, as reviewed by \citet{Land92} and
\citet{Schm01}.

From such measurements it is now known that the fields of white dwarfs
range in strength from some tens of kG up to hundreds of MG, with the
majority having fields in the range of $1 - 10^2$~MG. These fields are
apparently roughly dipolar in structure. The measured field may vary
periodically with time as the star rotates (although many white dwarf
fields show no sign at all of rotational variations), but it appears
that the fields change intrisically only on time scales that are too
long to have been detected so far.

The least studied regime of white dwarf magnetism is the low-field
end. This is largely due to the difficulty of detecting kG fields in
such faint (usually Johnson $V$ magnitude of 12 or fainter) objects.
As a result, only a very small number of magnetic fields below 100~kG
have been securely detected \citep{Aznaetal04,Kawketal12,Landetal12}.
Even fewer have been modelled in any useful way, so we really do not
yet even know if the weak fields found in a few white dwarfs are
similar in structure to the common stronger fields or if they
represent a significantly different phenomenon.

This deficiency in our knowledge of the weak field end of the field
strength distribution is important because we do not clearly
understand at present how the fields of white dwarfs arise. It has
been proposed that they may be the fossil remmnants of magnetism
inherited from an earlier stage of evolution, for example from the
main sequence \citep{Land92}. Alternatively, the fields of white
dwarfs may have a more specific origin, such as being produced during
the mergers of close binary systems
\citep{ValyFabr99,Toutetal08,Wicketal14}.

There are not many kinds of information about white dwarf magnetism
that really provide useful tests and constraints on theoretical ideas
about how magnetism arises in white dwarfs. The distribution of
magnetic fields strengths over the white dwarf population is one kind
of valuable information, but the low-field end of this distribution is
very poorly known. Another kind of constraint is provided by detailed
models of individual white dwarf magnetic fields. Modelling has proved
rather difficult at the high-field end of the distribution, but models
of field structure in the 1--30~MG range are fairly well developed
\citep{Wick01}.  At the lowest field, the lack of suitable candidate
stars to model as well as the lack of detailed and repeated observations of
the tiny number known of really weak-field white dwarfs has meant that
at present it is really not even clear whether the weakest magnetic
fields are similar in structure to larger fields in other magnetic
white dwarfs.

We have started a project to try to obtain the necessary observational
data to clarify both the frequency of very weak magnetic fields in white
dwarfs and the structure of fields found. Because the fields we seek
are at the limit of detectability, we have looked for new ways to
improve the sensitivity of our measurements. In this work, we describe
a novel technique to search for and study the weakest DA white dwarf
fields. We describe the new technique in detail, and test it on the
bright DA white dwarf \eri. We also report several conventional but
still very precise measurements of the field of \eri\ made to
corroborate measurements obtained with the new method, and to
allow a direct comparison of the field sensitivity limits obtained
with various methods. These measurements confirm the exceptional field
sensitivity of our new method, and the non-detection of a field
in \eri.

\section{A new technique for weak field detection in white dwarfs}

Detection of magnetic fields in white dwarfs is done using various
aspects of the Zeeman effect (and its large-field extensions). In the
presence of a field, single spectral lines split into multiple
components, and this splitting can be observed directly in simple
intensity (Stokes $I$) spectra if it is large enough. Measuring the
separation of the outer components provides a measurement of the mean
field modulus \bs, the magnitude of the field averaged over the
visible stellar hemisphere.

In addition, the component(s) of the split line that shift
systematically to longer and shorter wavelengths from the unperturbed
wavelength (the $\sigma$ components) become circularly polarised in
opposite senses if the field has a net component along the line of
sight. This causes the mean position of the spectral line to be at
slightly different wavelengths as observed in right and left
circularly polarised light.  This slight difference between the mean
positions of a spectral line as seen in the two polarisations can
conveniently be measured by forming the difference between the two
polarised spectra (Stokes $V$), as will be discussed below. Measuring
this $V$ difference spectrum makes possible the deduction of the mean
line-of-sight component of the field \bz.

Because almost all white dwarfs are quite faint (generally fainter
than magnitude 12 or 13), most white dwarf spectroscopy is done at
rather low resolving power $R$, typically 1000--2000, and often
with fairly low signal-to-noise ratio, of the order of 20 or 30. Under
these conditions, Zeeman splitting can be detected in the unresolved
cores of the broad (roughly 1--200~\AA\ wide) Balmer lines if a field
of at least several hundred kG is present. Most of the $\sim10^3$
magnetic white dwarfs now known have been identified in this manner
\citep[e.g.][]{Kepletal13}.

However, it is well known that the H$\alpha$ line in many DA white
dwarfs possesses a deep and narrow non-LTE line core.  Observations of
white dwarfs at high spectral resolution, for example in the context
of the ESO SPY search for SN~Ia progenitors \citep{Koesetal01,
  Koesetal09} can resolve this narrow feature.  Because the full width
of this line core is only about 1.5~\AA, much smaller line splitting
can be detected than is visible at low resolution, and fields as small
as about 20 or 30~kG can be identified and measured. The very
weak-field magnetic white dwarf WD2105-820 \citep{Landetal12} was
first identified as a candidate magnetic white dwarf by
\citet{Koesetal98} on the basis of slight broadening of the H$\alpha$
core by an amount which suggested the presence of a field with 
$\bs \sim 40$~kG.

To search for still weaker magnetic fields, spectropolarimetry is
essential. The small difference in radial velocity between spectral
lines as seen in the two senses of polarisation can be measured by
detecting the tiny differences in line wing depth of opposite sign in
the two line wings. This measurement is usually made with a low
resolution spectropolarimeter. By combining the signals from several
Balmer lines (or He lines), uncertainties of \bz\ can be as low as a
few hundred G, making possible the detection of longitudinal fields as
small as a few kG. Fields of $\bz \sim 3 - 10$~kG have been discovered
and measured with low-resolution circular spectropolarimetry using
FORS1 at ESO \citep{Aznaetal04,Landetal12}.

Several very efficient high-resolution spectropolarimeters are now
available on intermediate aperture telescopes (ESPeDOnS on the CFHT,
Narval on the TBL at Pic-du-Midi, and HARPSpol on the La Silla 3.6-m
telescope. These instruments have never, so far as we are aware, been
used for magnetic field measurement of white dwarfs. An important
reason that such instruments have not been used for white dwarf field
measurements is that all these instruments include cross-dispersed
echelle spectrographs with rather short orders, and are fibre-fed from
fibres that originate {\it after} the polarisation optics.  Any
broad-band polarisation signal is likely to be contaminated by
instrumental polarisation due to seeing and guiding fluctuations on
the entrance aperture, and to flexures in the Cassegrain polarimetric
unit, and to transmission variations in the fibres, as the telescope
moves. For these reasons, any broadband polarimetric signal present in
the light is usually set to zero during reduction. Polarisation is
normally measured with such instruments across spectral lines that are
very narrow compared to the length of an order, relative to an assumed
unpolarised continuum.  As a consequence, the tiny broad-band
circular polarisation (comparable in scale to whole orders) that may
be present in the broad wings of H lines in white dwarfs cannot
generally be measured reliably with high-resolution
spectropolarimeters.

\begin{figure}[ht]
\scalebox{0.37}{\includegraphics*{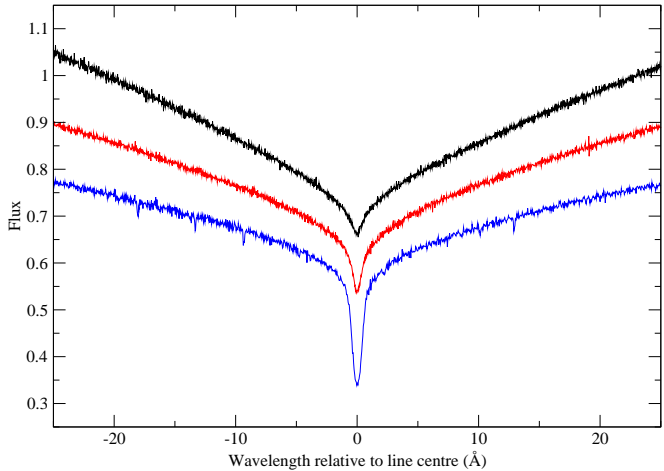}}
\caption{\label{Fig_comp_lines} Balmer lines H$\alpha$ (bottom),
  H$\beta$ (middle), H$\gamma$ (top) from a typical ESPaDOnS spectrum
  of \eri.. The continua for all lines are normalised to 1.0 in the
  far wings. The upper two spectra have been shifted upward for
  clarity by 0.2 (H$\beta$) and 0.3 (H$\gamma$). The unique shape of
  the core of H$\alpha$ is clear from this figure.  }
\end{figure}

However, DA white dwarfs with \te\ in the range $20000 \geq \te \geq
8000$~K do possess one spectral feature that is well suited to
polarimetric measurement with such high-resolution instruments: the
sharp and deep non-LTE core of H$\alpha$. This feature is illustrated
in Figure~\ref{Fig_comp_lines}, which compares the core profiles of
H$\alpha$, H$\beta$ and H$\gamma$ of the DA3 white dwarf \eri\ as
observed with a resolving power of $R = 65\,000$ (see below). The
uniquely sharp and deep nature of the core of H$\alpha$ is clear.
H$\beta$ also has a sharp core, but this core is substantially
shallower than that of H$\alpha$ and so the amplitude of a circular
polarisation signal from the H$\beta$ core would be several times
smaller than that of H$\alpha$. Note also that the sharp core of
H$\alpha$ weakens above $\te \sim 20\,000$~K, and disappears around
30\,000~K.

It is of course not obvious that a single spectral feature can provide
a useful level of field measurement precision, since the polarisation
measurement uses only about 2~\AA\ of the entire spectrum, while a
normal white dwarf \bz\ measurement uses nearly the full width of
several Balmer lines, and thus exploits up to half the stellar flux
observed with the instrument. Clearly the loss of information by using
a tiny fraction of the flux is severe. To compensate, however, the
slope of the line wings of the line core may be 20 or 30 times larger
than the slope of the broad Balmer line wings in a white dwarf.
Because of this, the polarisation signal in the wings of the non-LTE
line core is 20 or 30 times larger than the signal in the extended
line wings, and thus a measurement with ${\rm S/N} \sim 30$ times
lower (and a flux $\sim 10^3$ times smaller) than is required to
detect the broad line wing polarisation may make possible an accurate
\bz\ measurement using a high-resolution spectrograph.

We conclude that measurement of \bz\ using the core of H$\alpha$
might provide a substantial improvement in field measurement
uncertainty relative to measurements made by a conventional method. It
is clearly important to determine experimentally whether this novel
field measurement technique is actually sensitive enough to compete
with conventional low-resolution, full-spectrum spectropolarimetry. 

The one published set of white dwarf observations employing the
non-LTE core of H$\alpha$ that we know of was a series of measurements
in 1995 of the field of \eri\ using the polarimeter on the high
resolution Main Stellar Spectrograph at the Special Astrophysical
Observatory \citep{Fabretal03}. The authors report detection of a
variable field of semi-amplitude 2.3~kG varying with a period of a few
hours.  The reported measurements, which lasted 11~min each, had
uncertainties of typically 1.5 to 2~kG, so that the reported field
could only be detected by combining the individual meaurements.
However, the general potential of using the H$\alpha$ core to study
white dwarf fields was not clearly recognised, and this measurement
technique was not pursued further.

Modern spectropolarimeters have substantially higher efficiency than
the SAO spectrograph did in the 1990s.  Comparison of the shape of the
non-LTE line core of a few white dwarfs with the shape of individual
lines in rotionally broadened magnetic main sequence A and B stars
that have been previously measured with high resolution suggested that
an uncertainty of the order of $10^2$~G could be obtained for a white
dwarf of magnitude 9 in one hour. Clearly an experiment is called for
to discover what level of precision in \bz\ can be achieved in
practice for DA white dwarfs with a modern high-resolution
spectrograph.

\section{ESPaDOnS observations and data reduction}

We have tested this method of white dwarf field measurement using the
high-resolution spectropolarimeter ESPaDOnS at the
Canada-France-Hawaii Telescope, located on Mauna Kea, Hawaii. This
instrument uses a set of three Fresnel rhombs to make possible
measurement of all four Stokes parameters. The polarisation analyser,
mounted on the telescope) produces a double beam that is fibre-fed
into the entrance aperture of a cross-dispersed spectrograph having $R
\approx 65\,000$. The output spectra are recorded on a single CCD
chip, and subsequently reduced using the Libre-Esprit reduction
programme.

We have tested the potential of high-resolution H$\alpha$
spectropolarimetry for field measurement by observing \eri, the
brightest easily observed white dwarf in the sky. \eri\ is a DA3 star,
with $\te = 17100$~K and $\log g = 7.95$ \citep{Giametal12}. It is
thus a rather typical white dwarf with an H-rich atmosphere, except
for its particularly close distance of about 5 pc. Furthermore, as
noted above, a variable magnetic field of $\bz \sim 2$~kG was been
reported in the star by \citet{Fabretal03}.

Because \citet{Fabretal03} reported a rotation period of about 5
hours, we used 1-hour observing sequences (one sequence = four
observations with alternating waveplate positions). Such long
measurements would smear the \bz\ measurements over only about 0.2 cycle,
which should allow easy detection of the 2~kG field reported by the
SAO team if our precision goal was met. Two one-hour long sequences of
observations were made on each of three different nights, one hour
apart.

Each set of four subexposures produces an $I(\lambda)$ and a
$V(\lambda)$ spectrum covering approximately 3800~\AA\ to 1.04~$\mu$m
in about 40 orders. The reduction process also creates two different
null spectra and a table of pixel-by-pixel uncertainties $\sigma_i$.
H$\alpha$ is present in two orders (as is H$\beta$). We find that a
sharp non-LTE line core is also present in much
weaker form in H$\beta$, but completely absent from higher Balmer
lines. Thus almost all the signal worth working with is in two orders,
each of which covers the core and a part of the broad wings of
H$\alpha$.

\begin{figure}[ht]
\scalebox{0.37}{\includegraphics*{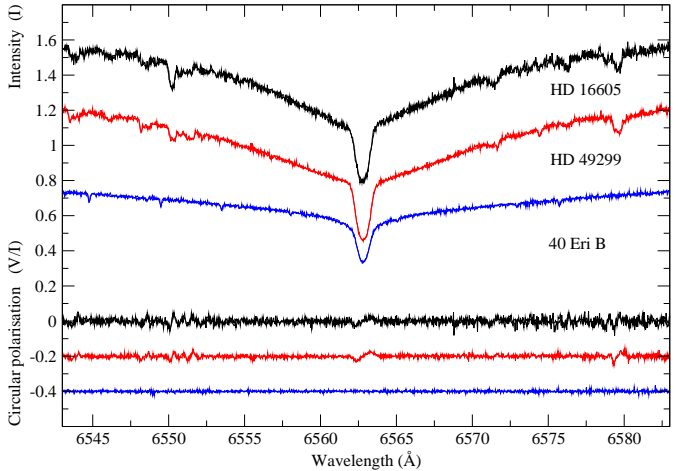}}
\caption{\label{Fig_esp_spec} The deep $I$ H$\alpha$ line core and a
  typical observed $V/I$ circular polarisation spectrum of \eri\ is
  compared to similar line cores and polarisation spectra of two
  magnetic Ap stars, HD~16605 and HD~49299, all observed with
  ESPaDOnS. The two magnetic Ap stars both have magnetic fields of
  $\bz \sim 2.5$~kG at the time of observation. The $V/I$ spectrum of
  \eri\ shows no sign of a magnetic signature at a level several times
  smaller than the signals clearly visible in the $V/I$ spectra of the
  Ap stars. All $I$ spectra are normalised to 1.0 in the continuum,
  shifted vertically by +0.3 (HD~42999) or +0.6 (HD~16605); $V/I$
  spectra are shifted downward from 0.0 by --0.2 (HD~49299) or --0.4
  (\eri). }
\end{figure}

The value of \bz\ is evaluated in these two orders (and in the two
orders that span the core of H$\beta$) with a small FORTRAN~90
programme that evaluates numerically the usual integral
\citep{Donaetal97},
\begin{equation}
   \bz = -2.14\,10^{12} \frac{\int v V(v) dv}
                            {\lambda_0 c \int (I_{\rm cont} - I(v))dv}
\end{equation}
where the intensity $I$ and circular polarisation $V$ are expressed as
function of velocity $v$ relative to the central wavelength of the
line $\lambda_0$, and $c$ is the speed of light \citep{Donaetal97}.
The Land\'{e} factor is assumed to be 1.0 for the Balmer lines. This
expression is exactly equivalent to evaluating the separation between
the centroids of the right and left circularly polarised line core
profiles, and $I_{\rm cont}$ is simply the upper boundary of the line
cores used in evaluating the centroid positions, so its level is
chosen to provide the highest possible S/N for the separation
measurement. 

With a discretely sampled spectrum, the integral is conveniently
approximated by a summation over the line:
\begin{equation}
   \bz \approx -2.14\,10^{12} \frac{\sum_{\rm line} v_i V_i}
                             {\lambda_0 c \sum_{\rm line} (I_{\rm cont} - I_i)}.
\end{equation}
where the summation extends over the pixels in the spectral line.
Examining the equation in summation form, it is clear, from the
general expression for the uncertainty of a sum of terms $i$ with
known Gaussian errors, that the uncertainty of \bz\ is given by
\begin{equation}
   \sigma_{\bz} \approx -2.14\,10^{12}
                        \frac{[\sum_{\rm line}(v_i \sigma_i)^2]^{1/2}}
                             {\lambda_0 c \sum_{\rm line} (I_{\rm cont} - I_i)},
\end{equation}
where $\sigma_i$ is the uncertaintiy assigned to $V_i$ by the
Libre-Esprit reduction programme. 

The integration window used basically coincides with the line core,
and was chosen (after some experimentation) to be 1.6~\AA\ in full
width.  The results from the two orders (which represent simultaneous
and essentially independent measurements of the field) are averaged.
The (averaged) field strength \nz\ resulting from one of the null
spectra (which should be zero within the uncertainty) was also
computed as a check on the reasonableness of the measurement and
particularly of the computed uncertainty $\sigma_{\bz}$.

Figure~\ref{Fig_esp_spec} shows that the polarisation signature that
we are searching for occurs in the line core of H$\alpha$. We have
tested our field measurement method further by using it to detrmine
the value of \bz\ from ESPaDOnS polarised spectra of several magnetic
Ap and Bp stars. In fact, the line cores of the Balmer lines in
magnetic Ap/Bp stars of low \vsi\ are quite similar to those of white
dwarfs: H$\alpha$ has a narrow and deep core, H$\beta$ has a shallower
sharp core, and the sharp cores quickly vanish as one moves up the
Balmer series. We have measured \bz\ in spectra of four such stars
using H$\alpha$, and compare it to the field values measured from the
same spectra using Least Squared Deconvolution (LSD) as described by
\citet{Landetal08}.

\begin{table}[th]
\begin{center}
\caption{\bz\ measured in magnetic sharp-line Ap/Bp stars with our
  H$\alpha$ method, compared to \bz\ measured from the same spectra
  using Least Squares Deconvolution}
\label{Tab_Bz_in_Aps}
\begin{tabular}{ccccccc }\hline 
   Star    &  MJD      & H$\alpha$ \bz    & LSD \bz    \\
           &           &    (Gauss)       &  (Gauss) \\
\hline
HD 16605   & 53570.58  & $-1380 \pm 135$  & $-2205 \pm 29$ \\
HD 49299   & 56293.42  & $-1935 \pm 74$   & $-2530 \pm 49$ \\
HD 133652  & 56437.35  & $+842 \pm 72$    & $+1330 \pm 18$ \\
HD 318107  & 54552.62  & $+4340 \pm 110$  & $+4900 \pm 45$ \\
\hline
\end{tabular}
\end{center}
\end{table}

The results of this comparison are shown in Table~\ref{Tab_Bz_in_Aps}.
Several conclusions may be drawn from these results. First, it is
clear that \bz\ measurements using the core of H$\alpha$ are 
practical and yield essentially the expected results. Secondly, the
\bz\ values resulting from the H$\alpha$ core are somewhat different
from those obtained with LSD. This is a very common situation; each
specific instrumental system used for measuring \bz\ yields, in
general, mildly different results than those obtained from other
systems \citep[see the discussion in Sec. 4 of][]{Landetal14}. Finally,
it appears that the field as measured using H$\alpha$ may be slightly
weaker than from other methods, as this is the case for all four
measurements in the table. However, no test has been possible yet to
directly compare \bz\ measurements obtained using the core of H$\alpha$
with those obtained by observing the wider Balmer line wings at low
resolution. We will assume until such tests are carried out that the
sensitivity of this new method is essentially the same as other
methods.  

The particulars of the observations of \eri\ and the resulting field
strength \bz, null field \nz, and uncertainty $\sigma$ are shown in
Table~\ref{table1}. Each observation lasted (with readout) 1.0~hr. \bz\ 
was evaluated for H$\beta$ as well as H$\alpha$, but the
corresponding uncertainties, around 450--600~G, are so large that we
did not combine them with the H$\alpha$ data or present them here.
Within their uncertainties, the H$\beta$ results are fully consistent
with the H$\alpha$ measurements.

\begin{table}[ht]
\begin{center}
\caption{Magnetic field measurements of \eri\ made with ESPaDOnS at
  the Canada-France-Hawaii Telescope. }
\label{table1}
\begin{tabular}{ccccc}\hline 
   Date    &  MJD      & \bz          & \nz \\ 
yr-mo-day  &           &    (Gauss)   &    (Gauss)  \\
\hline
2014-09-12 & 56912.502 & $174 \pm 91$ & $41 \pm 91$ \\
           & 56912.543 & $-14 \pm 88$ & $-91 \pm 88$ \\
2014-09-14 & 56914.579 & $-90 \pm 83$ & $-9 \pm 83$ \\
           & 56914.619 & $-44 \pm 83$ & $107 \pm 83$ \\
2014-09-16 & 56916.576 & $212 \pm 88$ & $147 \pm 88$ \\
           & 56916.617 &  $77 \pm 91$ & $26 \pm 91$ \\
\hline
\end{tabular}
\end{center}
\end{table}

The essential information derived from these spectra is shown in
Figure~\ref{Fig_esp_spec}. This figure displays a single $I$ spectrum
of \eri, and the corresponding $V/I$ spectrum (all the $I$ and $V/I$
spectra are essentially identical). For comparison, we also display
the $I$ and $V/I$ ESPaDOnS spectra of the A0p -- A1p main sequence
stars HD~16605 and HD~49299, which at the times of observation both
had magnetic fields of $\bz \approx -2.5$~kG. In spite of being main
sequence stars, these stars have H$\alpha$ line cores that are rather
similar to that of \eri, although the broad line wings of the Ap star
clearly have larger slopes than those of the white dwarf. The spectra
of the two main sequence stars illustrate how the $V/I$ H$\alpha$ line
core spectrum of a magnetic star with a kG field would appear, and
also shows clearly how the line core signal is enormously strengthened
relative to the (invisibly small) signal in the broad H$\alpha$ line
wings by the sharpness of the narrow core.

It is clear from the figure and from Table~\ref{table1} that the $V/I$
spectrum of \eri\ shows no significant sign of a magnetic signature,
and that any field present must be at least a factor of five or so
smaller than that of the Ap stars, in agreement with the results of
the numerical evaluation of \bz. We conclude that the ESPaDOnS field
measurements put a probable limit on any slowly-varying (time scale of
hours or more) field \bz\ in \eri\ of roughly 250~G, at least at the
times when we observed.

It is important to consider the limitations of such measurements that
may be introduced by the rotation of a white dwarf. Because white
dwarfs are small (with radii of the order of $10^4$~km) the normal
upper limits derived for projected rotation velocity, typically about 
30~\kms, lead to lower limits on rotation periods of the order of half
an hour \citep{Hebeetal97}. Furthermore, a few strongly magnetic white
dwarfs are actually observed to have rotation periods of less than an hour
\citep{Brinetal13}. Thus a measurement such as ours, lasting an hour
or more, might in some cases last longer than the rotation period of
the star. If the \bz\ field measurement reverses sign during rotation,
the measured field could average out to a substantially smaller value
than would be observed by measurments lasting only a few minutes.

This probably not a major measurement limitation in the specific case
of our measurements of \eri. \citet{Hebeetal97} have shown that the
rotation period of \eri\ is longer than about 2~hr, so that each
single measurement of the star is at most about half a rotation. Thus
some of our measurements at least should have observed mainly one magnetic
hemisphere of any dipole-like field, and should have detected a field
with about the sensitivity given in Table~\ref{table1}.  However, this
potential limitation of white dwarf field measurements based on long
integrations should be kept in mind.

It is worthwhile noting one further important strength of the
high-resolution H$\alpha$ \bz\ measurement. As already pointed out by
\citet{Fabretal03}, observation with a high-resolution spectrograph
has the advantage that the $I$ profile of the non-LTE core of
H$\alpha$ can also be studied, and thus the value (or an upper limit)
of \bs, measured at the {\it same time} as the \bz\ measurement, can be
obtained from the observations. The resulting measurement pairs can be
very useful for initial field structure modelling.

\section{Field measurements with the SAO Main Stellar Spectrograph}

To test the correctness of the high-resolution H$\alpha$ emasurements,
and for a comparison of conventional measurement methods of white dwarf
field strength with measurement using H$\alpha$ core, we report
measurements of the magnetic field of \eri\ made using two
lower-resolution spectropolarimeters, the MSS at the Special
Astrophysical Observatory, and ISIS at the William Herschel Telescope.

Measurements of Zeeman polarisation of \eri\ were carried out with the
main stellar spectrograph (MSS), permanently installed at the 6-m
telescope (BTA) at the Special Astrophysical Observatory of the
Russian Academy of Sciences, located on Mt. Pastukhova near
Zelenchukskaya in the Karachai-Cherkessian Republic, Russia. The MSS
is a longslit spectrograph equipped with a circular polarisation
analyser combined with an image slicer \citep{Chou04}.  The analyser
has a rotatable quarter-wave plate that is able to take two fixed
positions corresponding to the angles $-45\deg$ and $+45\deg$ relative
to the principal axes of the birefringent crystal. More details about
the instrument and observational technique can be found in
\citet{Joshetal12} and \citet{Kudretal06}. The MSS spectropolarimeter
exploits two regimes of spectral resolution: $R = 15000$ and $R =
6000$. We used both of the regimes in observations of \eri. In
contrast to the previous observations of this white dwarf with this
instrument \citep{FabrValy99}, the image slicer now employed provides
the MSS with much higher stability, which makes it possible to carry
out measurement of stellar longitudinal fields with uncertainties as
low as 30~G \citep{Kudretal06}.

\begin{figure}[ht]
\scalebox{0.33}{\includegraphics*[angle=-90]{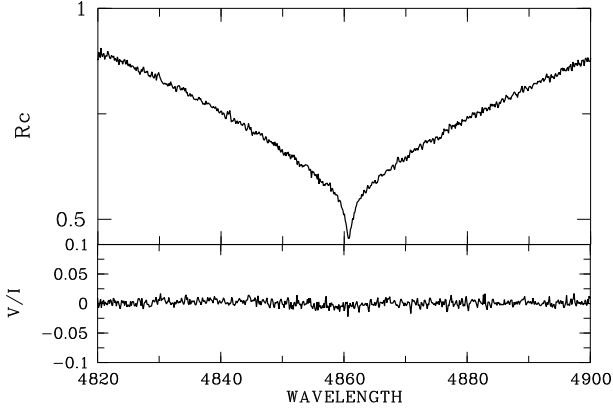}}
\caption{\label{Fig_sao_spec} The intensity ($I$) and circular
  polarisation ($V/I$) spectra of the H$\beta$ line observed with the
  MSS at SAO with $R = 15000$. The weak sharp core of H$\beta$ (also
  illustrated in Figure~\ref{Fig_comp_lines}) is clearly resolved in
  these data. No sign of any magnetic polarisation signature is seen
  at line centre.  }
\end{figure}

All observations of \eri\ were conducted following the standard
procedures that we usually use \citep[for instance][]{Valyetal05}. A
polarisation observation consists of a series of paired exposures
obtained at two orthogonal orientations of the quarter-waveplate. The
data reduction is also standard.

We usually make longitudinal field measurements of white dwarfs by
measurement of the displacement between the positions of spectral
lines in the spectra of opposite circular polarisations. For
determination of the line profile displacement we used a technique
based on estimates of the autocorrelation function of Zeeman shifted
spectral features using Fourier transforms \citep{Scar89}. This
technique is insensitive to uncertainties in the continuum
determination \citep{Klocetal96}, a feature that is important in
observations with the MSS.  Comparison of the sensitivity of this
method with other methods \citep{Monietal02} and our own experience
make this method preferable in low or moderate resolution observations
of magnetic white dwarfs with the MSS.  However, the high-resolution
(R15000) observations of 40EriB reported here, when reduced with this
method, were found to exhibit artificially small error bars. Due to
this problem the high-resolution estimate of the field of \eri\
measured with R15000 was also made by the regression method, which
demonstrated much more robust accuracy in measurements of the narrow
central H$\beta$ profile. This particular observation can be
compared fairly directly with the data from the CFHT.

The reduced magnetic field measurements of \eri\ obtained with the MSS
are presented in Table~2. The columns are self-explanatory. The first
observation, using only H$\beta$, was obtained with $R = 15000$, while
the rest were obtained at $R = 6000$. Figure~\ref{Fig_sao_spec} shows
the $I$ and $V/I$ spectra obtained with the R15000 resolution. The
sharp core is clearly visible at this resolution. However, the fact
that the core of H$\alpha$ is about three times deeper than that of
H$\beta$, and slightly narrower, means that the S/N is H$\alpha$ is
about four times higher than that of H$\beta$ for \bz\ measurements
made with both lines, as found with the ESPaDOnS data described above.

\begin{table}[h]
\begin{center}
\caption{Magnetic field measurements of \eri\ made with the Main
  Stellar Spectrograph at the Special Astrophysical Observatory. }
\label{table2}
\begin{tabular}{ccccccc }\hline 
   Date    &  MJD    & $t_{\rm exp}$ & \bz          & Lines \\
yr-mo-day  &           &   (sec) &    (Gauss)       &  used  \\
\hline
2014-10-09 & 56938.944 & 3600    & $+573 \pm 290$   & H$\beta$
\\
2014-10-10 & 56940.042 & 7200    & $-46 \pm 638$    & H$\beta$
-- H$\delta$ \\
2014-10-11 & 56941.021 & 9600    & $-208 \pm 434$   & H$\beta$
-- H$\delta$ \\
2014-10-12 & 56942.000 & 7200    & $898 \pm 569$    & H$\beta$
-- H$\delta$ \\
2014-10-14 & 56944.000 & 3600    & $-43 \pm 780$    & H$\beta$
-- H$\delta$ \\
\hline
\end{tabular}
\end{center}
\end{table}

These observations were all taken within a span of five nights. It is
possible that \eri\ rotates slowly enough that these measurements all
sample one hemisphere. In this case it is meaningful to average all
the field measurements together. The mean field measured using all
five measurements (R6000 and R15000 together) is $\bz = +340 \pm
202$~G. Averaging only the four R6000 data together yields $\bz = +119
\pm 283$~G. 

None of these observations, taken individually or together, provide
strong evidence for the presence of a field. From these measurements,
it does not appear that the \bz\ field of \eri\ observed by us exceeds
perhaps 1~kG. This is fully consistent with the measurements made with
ESPaDOnS, but is inconsistent with the earlier report of a field of
$\bz \sim 3 - 5$~kG varying sinusoidally with a period of about
5~hours \citep{Fabretal03}. If a field of that strength were present
in \eri, we would expect that we should have detected it at the level
of $5 - 6 \sigma$ in at least two or three of the recent observations.
It now appears that the old SP-124 spectropolarimeter used for most of
the earlier observations of \eri\ had larger uncertainties than was
recognised at the time, and so we no longer consider the field
reported by \citet{Fabretal03} to be securely detected.  This problem
has already been briefly discussed in the context of observation of
the weak-field white dwarf WD0009+501 \citep{Valyetal05}

\section{Field measurements with ISIS at the WHT}

\begin{figure*}[ht]
\scalebox{0.60}{\includegraphics*[angle=-90]{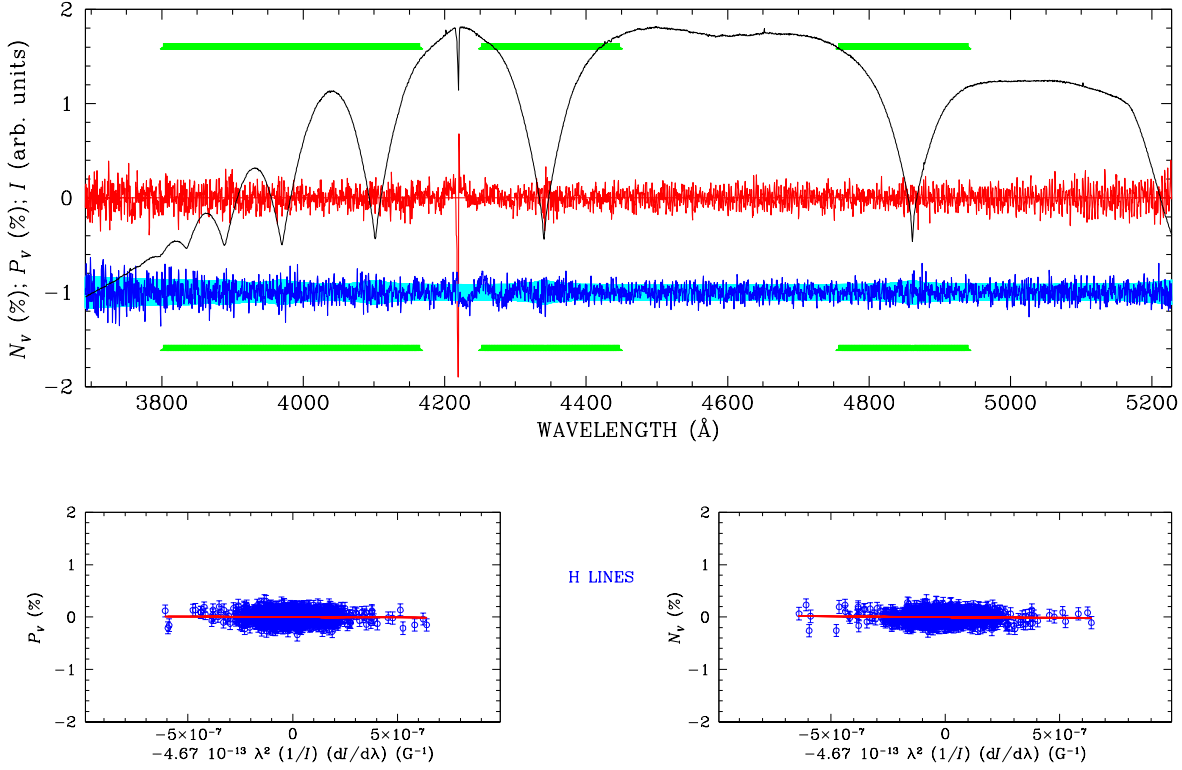}}
\caption{\label{Fig_isis_spec} Observation of \eri\ obtained with ISIS
  on MJD~57056.92. The details shown in the figure are explain ed in
  the text. The \bz\ and \nz\ values deduced from the correlation
  diagrams in the lower part of the figure are $158 \pm 204$~G and
  $-254 \pm 186$~G, respectively.  }
\end{figure*}

Magnetic observations of \eri\ were carried out during a mission
with the ISIS (Intermediate dispersion Spectrograph and Imaging
System) at the William Herschel Telescope at the Observatorio del Roque
de los Muchachos on La Palma in the Canaries. This is an
intermediate resolution spectrograph that can be used as a
spectropolarimeter with the addition of a rotatable quarter-wave plate
above the slit, followed by a Savart plate (a linearly polarising
beam-splitter) immediately below the slit. The wave-plate axis is set
alternately at $+45\deg$ and $-45\deg$ to the principal axes of the
Savart plate. 

Observations were carried out using the R600B grating, covering a
useful spectral range of 3600 to 5200~\AA. The slit was set to
1~arcsec, leading to a slit image width of about 1.8~\AA, and a
resolving power of $R \approx 2450$. With these instrument settings
the weak sharp core of H$\beta$ is not resolved.

During the three-night run there were frequent thin clouds. The
seeing was quite variable, between 0.4 and 2~arcsec, and integration
times were adjusted on the fly to compensate. 

The three observations were reduced using the same regression
technique used to reduce spectropolarimetric observations made with
the (rather similar) FORS spectropolarimeter at ESO. A correlation
plot is created between the local slope of the $I$ profile and the
local value of $V/I$ in the circular polarisation spectrum, and the
magnetic field is deduced from slope of the best fit correlation line
using the expression 
\begin{equation}
   V(\lambda) = -g_{\rm eff} C_Z \lambda^2
         \frac{{\rm d}I(\lambda)}{{\rm d}\lambda} \bz    
\end{equation}
where $C_Z = e/4 \pi m_{\rm e} c^2 = 4.67\,10^{-13}$~\AA$^{-1}$G$^{-1}$.
This method is discussed in considerable detail by
\citet{Bagnetal02,Bagnetal12}.  

The reduction technique is illustrated in Figure~\ref{Fig_isis_spec}.
In the upper panel, the uncalibrated stellar spectrum is shown in
black in arbitrary units. At the top and bottom of this panel, green
bands indicate the regions of spectral lines used for the reduction.
The red noise band centred on 0 is the observed $V/I = P_V$ spectrum.
The blue band below centred on --1 is a null spectrum $N_V$ which
should show no Zeeman polarisation signal, but should have the same
noise behaviour as $P_V$. Within the null spectrum, a continuous light
blue band indicates the uncertainty of each pixel in the spectra. 

The two lower panels show the correlation diagrams from which the
field strength \bz\ is deduced, constructed by plotting all the point
pairs of $P_V$ against the corresponding local values of $-4.67\,10^{-13}
\lambda^2 (1/I) ({\rm d}I/{\rm d}\lambda)$. The left panel is the
correlation diagram for $P_V$, the right the corresponding diagram for
$N_V$. The horizontal lines are the fits to the correlation diagrams,
and in this case neither exhibits significant slope, indicating that
\bz\ and \nz\ are both zero within uncertainties. 

The results of the three observations are shown in Table~\ref{table3}.
None of the measurements of \bz\ is significantly different from
zero. With the uncertainties obtained, it again seems clear that a
field of about 2~kG should have been detected.

\begin{table}[h]
\begin{center}
\caption{Magnetic field measurements of \eri\ made with the ISIS
  medium-resolution spectrograph at the Willim Herschel Telescope}
\label{table3}
\begin{tabular}{ccccccc }\hline 
   Date    &  MJD    & $t_{\rm exp}$ & \bz          & \nz    \\
yr-mo-day  &           &   (sec) &    (Gauss)       &  (Gauss) \\
\hline
2015-02-02 & 57055.89  & 4000    & $199 \pm 393$    & $91 \pm 391$ \\
2015-02-03 & 57056.92  & 2800    & $158 \pm 204$    & $-254 \pm 186$ \\
2015-02-04 & 57057.95  & 3200    & $574 \pm 258$    & $184 \pm 261$ \\
\hline
\end{tabular}
\end{center}
\end{table}

\section{Comparison of methods}

The data sets above (and comparable data available from white dwarf
magnetic observations made with FORS1) make it possible to understand
which of the four spectropolarimeters are best adapted for searching
for the very weakest fields that may be present in bright DA white
dwarfs in the temperature range of 8000--20\,000~K. We can make a
useful comparison by computing the integration times that would be
needed with these instruments to reach a given uncertainty in \bz\ for
a given star. Since the brightest white dwarfs have $V$ magnitudes in
the range of 12--13, an interesting comparison is to use the data
reported above to estimate the integration times that would be
required with various instruments to obtain measurement uncertainties
of 500~G for a star like \eri\ (i.e. a star with a similar \te) which
has a $V$ magnitude of 12.5. 

This comparison is shown in Table~\ref{table4}. The values for
the three spectropolarimeters discussed above are scaled directly from
the most precise measurement of \eri\ reported for each instrument
(the measurement which reaches the highest ratio of S/N to the square
root of the integration time). We use the data summarised by
\citet{Landetal12} to include FORS1 in the table, using the best of
three measurements of WD2149+021, a white dwarf with almost exactly
the same \te\ as \eri.

\begin{table}[ht]
\begin{center}
  \caption{Comparison of integration times $t_{\rm int}$ required by
    each of the spectropolarimeters discussed above, and of ESO FORS,
    to reach $\sigma_{\bz} = 500$~G for an observation of a DA3 star
    of V = 12.5 }
\label{table4}
\begin{tabular}{lr }
\hline 
   Spectropolarimeter   &  $t_{\rm exp}$    \\
                        &    (sec)         \\
\hline
CFHT ESPaDOnS H$\alpha$               & 1500    \\
SAO MSS R15000 H$\beta$               & 19200   \\
WHT ISIS blue arm R600B grating       & 7400    \\
ESO FORS grism 600B, 3470--5890~\AA\  & 1950    \\
\hline
\end{tabular}
\end{center}
\end{table}

This table shows clearly that for the specific goal of detecting the
weakest possible magnetic field in DA3--DA6 type white dwarf, the best
choices are the high-resolution H$\alpha$ method and conventional
spectropolarimetry using FORS in the blue, with roughly similar
integration times. Amazingly, the measurement would apparently take
about the same amount of time with ESPaDOnS on the 3.6-m CFH Telescope
as with FORS on an 8-m telescope. An efficient high-resolution
spectropolarimeter on an 8-m telescope would require about 300~s of
sky time to carry out this standard comparison measurement.

ISIS has about the same global efficiency as FORS, but has only
about 1/4 the telescope mirror area, which increases the required
exposure times by about factor of four relative to a measurement with
FORS. For our measurements, the SAO MSS spectropolarimeter has not been
used in an optimal mode; the best measurement is the one made with
R15000 but the efficiency would probably have been considerably higher
with H$\alpha$. Efforts are being made to bring this mode into
convenient operation. In the meantime, the SAO MSS spectropolarimeter
is best used for measurements of stars in which the magnitude of \bz\
reaches at least 20--30~kG.

For hotter DA1--DA2 white dwarfs in which the deep H$\alpha$ core is
weak or absent, so that the H$\alpha$ method with a medium- or
high-resolution spectropolarimeter is not effective, the best choices
at present for the deepest field surveys are FORS in the southern sky
and ISIS in the north.

\section{Conclusions}

Known magnetic fields in white dwarf stars range in field strength
from a few kG to several hundred MG, an impressive five orders of
magnitude. This fact, and the more detailed distribution of magnetic
field strength and field structure, surely contain important clues
about the origin and evolution of the observed, usually fairly simple,
fossil fields.

Over this distribution of field strength, the most poorly studied
section is the weak field end. White dwarfs with \bz\ of
a few kG are rare (only three or four are known) and the known stars
have not been extensively studied. We are engaged in filling in gaps
in our knowledge of such stars. 

The biggest obstacles to finding and studying the weakest field white
dwarfs come from their general faintness (almost all fainter than $V =
12$) combined with the extraordinary breadth of the few spectral lines.
In the commonest white dwarfs, with H-rich atmospheres (DA stars) the
Balmer lines can be 200~\AA\ wide. Such broad lines yield only a very
weak polarisation signal when the field is a few kG in strength.

We have proposed studying the fields of DA stars in the temperature
range of at least 8\,000--20\,000~K using the sharp and deep line core
of H$\alpha$, which in main sequence stars yield clearly recognisable
polarisation signatures with fields of a few hundred~G. We have tested
the usefulness of this field measurement method by observing the
bright DA white dwarf \eri, in which a variable field of a few kG has
been reported \citep{Fabretal03}. The star was observed repeatedly
with the high-resolution ($R = 65\,000$) ESPaDOnS spectropolarimeter
at the Canada-France-Hawaii Telescope. Reduction of the data showed
that indeed field measurement using the core of H$\alpha$ is very
sensitive (measurements of \bz\ with uncertainties $\sigma_{\bz}$ as
low as 83~G were obtained in one-hour integrations), but no trace of a
magnetic field was found (Table~\ref{table1}). We conclude that \bz\
was almost certainly below 250~G at the times when we observed \eri. 

Measurements of the \bz\ values of \eri\ were also obtained with two
other spectropolarimeters, the MSS of the SAO, and ISIS at the WHT
(Table~\ref{table2} and Table~\ref{table3}, to validate the results
found with ESPaDOnS. These measurements confirm
the lack of a detectable field in \eri\ at the 2--3~kG level. The
observations also help to evaluate the important gain in field sensitivity
realised with the H$\alpha$ Zeeman measurements relative to the
sensitivity achieved with normal low-resolution spectropolarimeters.
The field measurements using H$\alpha$ reached uncertainties more than
a factor of two smaller than were obtained with the other instruments.

This comparison was extended by using the observational data from this
paper, together with published white dwarf field measurements made
with the ESO FORS low-resolution spectropolarimeter, to estimate the
integration time needed to make a measurement of \bz\ accurate to
$\sigma_{\bz} = 500$~G on a star similar to \eri, but of magnitude $V$ =
12.5.  It is found that the integration time required with ESPaDOnS
(on a 3.6-m telescope) using only H$\alpha$ is actually a little
shorter than the time needed with FORS (on an 8-m telescope). The
high-resolution measurement is clearly substantially more efficient than
low-resolution observations on telescopes of the 4-m class for white
dwarfs with suitable H$\alpha$ lines.

The next step is to use this new technique to survey the brightest
cool and intermediate DA white dwarfs of the northern hemisphere in
order to identify a few more of the extremely weak-field white dwarfs,
and to study the structure of their magnetic fields. 

\begin{acknowledgements}

  JDL acknowledges ongoing financial support from the Natural Sciences
  and Engineering Research Council of Canada.  DG, ES and GG
  acknowledge the Russian Scientific Foundation (grant N14-50-00043);
  GV acknowledges the Russian Foundation for Basic Research (RFBR
  grant N15-02-05183)

\end{acknowledgements}

\bibliography{MyBiblio}

\end{document}